# *dq* Modeling for Series-Parallel Compensated Wireless Power Transfer Systems

Zixuan Jiang
Jiangsu University
Jiangsu, China
3220501146@stmail.ujs.edu.cn

*Abstract*—Series-parallel (SP) compensated wireless power transfer (WPT) systems are widely used in some specific scenarios, such as bioelectronics and portable electronics. However, most studies are based on the phasor method and focused on the steady-state analysis, which may overlook the transient process of systems. Accordingly, inspired by the notion of coordinate transformation in the field of motor drive, this work develops a *dq* modeling method for SP compensated WPT systems. The proposed model effectively characterizes first-order system dynamics, facilitating enhanced measurement precision and control system development. One measurement application, *dq* model-based mutual inductance identification, is presented to reflect the value of the *dq* model. Simulation results are shown to validate the model's effectiveness, indicating that the developed model can be a good tool for the design of SP compensated WPT systems.

## I. Introduction

With the rapid evolution of diverse renewable energy technologies, energy transmission ways have undergone transformative diversification. In contrast to conventional wired charging, wireless power transfer (WPT), particularly inductive WPT, has emerged as a potential way, garnering great attention from both academia and industry due to its meaning advantages. These advantages include convenience, elimination of mechanical wear and connector aging, resilience in harsh environmental conditions, and seamless integration with smart devices. Such merits position WPT as a potential technique for some special electric-driven applications, such as unmanned aerial vehicles, electric vehicles, and bio-electronic devices [1]-[2]. To better understand and apply this technique, mathematical models are required to assist in system parameter selection and control algorithm design [3]. So far, existing research has primarily focused on two categories of mathematical models [4]-[9].

The first category for WPT comprises sinusoidal steady-state phasor-based models. In [4], hybrid topologies employing series-series (SS) and parallel-series (PS) or series-parallel (SP) and parallel-parallel (PP) compensation are proposed for battery charging. With the steady-state model, the authors design proper parameters, thus realizing constant-current (CC) and constant-voltage (CV) charging under the same transformer, compensation capacitors, and operating frequency. In [5], a mutual inductance dynamic prediction scheme is introduced by using steady-state circuit model of LCC-parallel compensated WPT systems. Then, a CC control scheme is designed for in-flight wireless charging systems. Pang et al. derive the phasor model for CLC-series compensated WPT systems, and then design an impedance buffer for it, which functions as either an inductor or capacitor to fully counteract reactance caused by inevitable parameter perturbations and active frequency adjustments [6]. Most of these control schemes lean on the steady-state traits of WPT systems. While static models provide theoretical benchmarks, they ignore the dynamic properties of systems, which may not reliable for high-performance application scenarios.

Another type of models employs time-domain differential equations to describe the system. Chow et al. [7] derives a detailed time-domain circuit model incorporating the nonlinear load characteristics of diode bridge rectifiers. With it, the value of mutual inductance can be estimated by solving fourth-order differential equations. In [8], an accurate time-domain model for the SS compensated WPT systems with a dc-dc converter is presented, which is conducive to the assessments and the accurate designs of the system. In [9], a novel dynamic modeling approach is introduced based on coupled-mode theory for magnetically resonant WPT systems. The slowly changed amplitudes and phases of coupled modes are used, which aims to provide an abundant insight into the dynamic behaviors of the resonators. However, time-domain models are inherently more complex, involving differential equations, state-space representations, and other computationally intensive formulations. This necessitates high-performance real-time computing hardware, particularly in systems requiring rapid responses, which limits the wide applications of the WPT technique.

In response to the aforementioned challenges, this paper focuses on a modeling approach. The main contributions are shown below. First, a detailed derivation of the *dq* model for SP compensated WPT systems is proposed, which is inspired by three-phase motor coordinate transformation [10] and the work in [11]. The SP compensation is selected as our research object due to its CV output trait [12] and good tolerance under coil misalignment [13]. The proposed *dq* model can depict the first-order dynamic traits of the system while preserving computational efficiency and implementation simplicity. Second, a mutual inductance identification scheme based on the model is designed to demonstrate its practical utility. Simulation results validate the effectiveness of the proposed model as well as the identification method.

## II. Modeling

### A. αβ Model

The circuit of the SP compensated WPT system is shown in Fig. 1(a), which constitutes as an *α*-axis model in the *αβ* coordinate system. Then, an imaginary circuit with a 90° phase lag than the *α*-axis model is constructed as a *β*-axis model, as shown in Fig. 1(b). Wherein, $L_t$ ($L_r$), $C_t$ ($C_r$), and $R_t$ ($R_r$) are the equivalent inductance, the capacitance, and the equivalent resistance of the transmitting (receiving) side, respectively. Additionally, $R_L$ is the equivalent load resistance, and $M$ is the mutual inductance. The input voltage $u_s$, the transmitting current $i_t$, the current $i_r$ through the $L_r$, and the current $i_c$ through the $C_r$ in the $\alpha$-axis model are defined as

$$\begin{cases} u_s = U_s \cos(\omega t) \\ i_t = I_t \cos(\omega t - \varphi_t) \\ i_r = I_r \cos(\omega t - \varphi_r) \\ i_c = I_c \cos(\omega t - \varphi_c) \end{cases} \quad (1)$$

where $\omega$ is the angular frequency, $U_s$, $I_t$, $I_r$, and $I_c$ are the amplitude values, and $\varphi_t$, $\varphi_r$, and $\varphi_c$ are the initial phase angles. The $\alpha\beta$ model shown in Fig. 1(c) can be obtained by using Euler's formula in the $\alpha\beta$ coordinate system, where the current vectors can be expressed as

$$\begin{cases} I_{\alpha\beta t} = I_t e^{j(\omega t - \varphi_t)} = I_{dqt} e^{j\omega t} \\ I_{\alpha\beta r} = I_r e^{j(\omega t - \varphi_r)} = I_{dqr} e^{j\omega t} \\ I_{\alpha\beta c} = I_c e^{j(\omega t - \varphi_c)} = I_{dqc} e^{j\omega t} \end{cases} \quad (2)$$

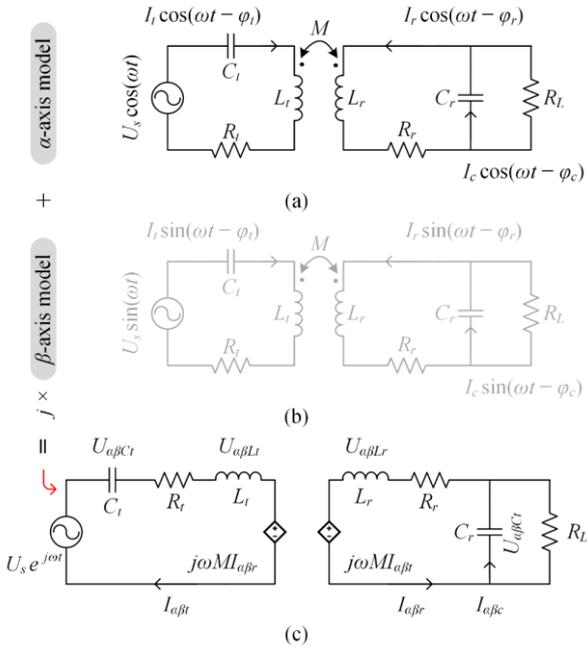

Fig. 1. Circuits. (a) Equivalent circuit of SP compensated WPT system. (b) Imaginary circuit. (c) Circuit of $\alpha\beta$ model.

Based on the work [11], the voltage vectors of the inductors are given as

$$U_{\alpha\beta Lt} = L_t \frac{dI_{\alpha\beta t}}{dt} = \left[ L_t \frac{dI_{dqt}}{dt} + j\omega L_t I_{dqt} \right] e^{j\omega t} \quad (3)$$

$$U_{\alpha\beta Lr} = L_r \frac{dI_{\alpha\beta r}}{dt} = \left[ L_r \frac{dI_{dqr}}{dt} + j\omega L_r I_{dqr} \right] e^{j\omega t} \quad (4)$$

Similarly, after ignoring the high-order terms, the voltage vectors across the capacitors can be presented as

$$U_{\alpha\beta Ct} = \frac{1}{C_t} \int I_{\alpha\beta t} dt = \left[ \frac{1}{\omega^2 C_t} \frac{dI_{dqt}}{dt} + \frac{1}{j\omega C_t} I_{dqt} \right] e^{j\omega t} \quad (5)$$

$$U_{\alpha\beta Cr} = \frac{1}{C_r} \int I_{\alpha\beta c} dt = \left[ \frac{1}{\omega^2 C_r} \frac{dI_{dqc}}{dt} + \frac{1}{j\omega C_r} I_{dqc} \right] e^{j\omega t} \quad (6)$$

With the derived voltage vectors and Kirchhoff's voltage law, the $\alpha\beta$ model, shown in Fig. 1(c), can be obtained as

$$U_s e^{j\omega t} = \left[ R_t I_{dqt} + \left( L_t + \frac{1}{\omega^2 C_t} \right) \frac{dI_{dqt}}{dt} \right. \\ \left. + j\left( \omega L_t - \frac{1}{\omega C_t} \right) I_{dqt} + j\omega M I_{dqr} \right] e^{j\omega t} \quad (7)$$

$$0 = \left[ R_r I_{dqr} + L_r \frac{dI_{dqr}}{dt} + j\omega L_r I_{dqr} \right. \\ \left. + j\omega M I_{dqt} + \left( I_{dqr} - I_{dqc} \right) R_L \right] e^{j\omega t} \quad (8)$$

$$0 = \left[ \frac{1}{\omega^2 C_r} \frac{dI_{dqc}}{dt} + \frac{1}{j\omega C_r} I_{dqc} - \left( I_{dqr} - I_{dqc} \right) R_L \right] e^{j\omega t} \quad (9)$$

Notedly, the currents and voltages shown above are all rotational vectors with a constant angular velocity $\omega$ in the $\alpha\beta$ coordinate system.

*B. dq Model*

Given that (7), (8), and (9) all have the same rotational factor $e^{j\omega t}$, a *dq* synchronous coordinate system with the same velocity is naturally introduced. As shown in Fig. 2, the rotational vectors can constitute as stationary vectors in the *dq* synchronous coordinate system, and the *dq* model is obtained by eliminating the same rotational factor on both sides of the equations.

$$U_{dqs} = R_t I_{dqt} + \left( L_t + \frac{1}{\omega^2 C_t} \right) \frac{dI_{dqt}}{dt} \\ j\left( \omega L_t - \frac{1}{\omega C_t} \right) I_{dqt} + j\omega M I_{dqr} \quad (10)$$

$$0 = R_r I_{dqr} + L_r \frac{dI_{dqr}}{dt} + j\omega L_r I_{dqr} \\ + j\omega M I_{dqt} + \left( I_{dqr} - I_{dqc} \right) R_L \quad (11)$$

$$0 = \frac{1}{\omega^2 C_r} \frac{dI_{dqc}}{dt} + \frac{1}{j\omega C_r} I_{dqc} - \left( I_{dqr} - I_{dqc} \right) R_L \quad (12)$$

where $U_{dqs}$ is a stationary voltage vector with zero phase angle and a magnitude of $U_s$. Fig. 3 shows the *dq* model of the SP compensated WPT system, which can reflect the first-order dynamic properties of the system. It can help us fulfill high-performance measurement or control.

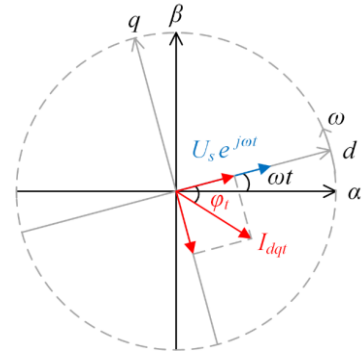

Fig. 2. Relationship between $\alpha\beta$ coordinate system and *dq* synchronous coordinate system.

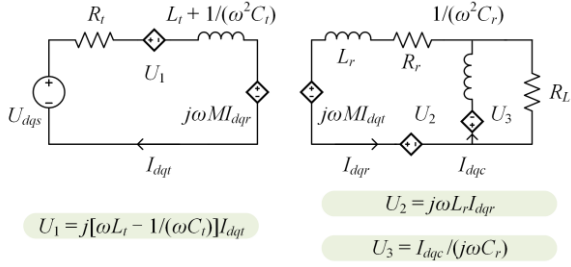

Fig. 3. *dq* model of SP compensated WPT system.

As a complex form model, the *dq* model can be partitioned into the real parts (*d*-axis circuits) and the imaginary parts (*q*-axis circuits). Fig. 4 shows the block diagrams of the *dq* model, where the receiving side is exemplified. Wherein, $I_{dqt}$, $I_{dqr}$, and $I_{dqc}$ can be decomposed as

$$\begin{cases} I_{dqt} = I_{dt} + jI_{qt} \\ I_{dqr} = I_{dr} + jI_{qr} \\ I_{dqc} = I_{dc} + jI_{qc} \end{cases} \quad (13)$$

Then, the relationships between the amplitudes of the currents and the magnitudes of the current vectors are expressed as

$$\begin{cases} I_t = |I_{dqt}| = \sqrt{I_{dt}^2 + I_{qt}^2} \\ I_r = |I_{dqr}| = \sqrt{I_{dr}^2 + I_{qr}^2} \\ I_c = |I_{dqc}| = \sqrt{I_{dc}^2 + I_{qc}^2} \end{cases} \quad (14)$$

Taken the transmitting side as an example, the relationship between the initial phase angle of the current and the magnitudes of the current vectors is expressed as

$$\varphi_t = -\arctan\left(\frac{I_{qt}}{I_{dt}}\right) \quad (15)$$

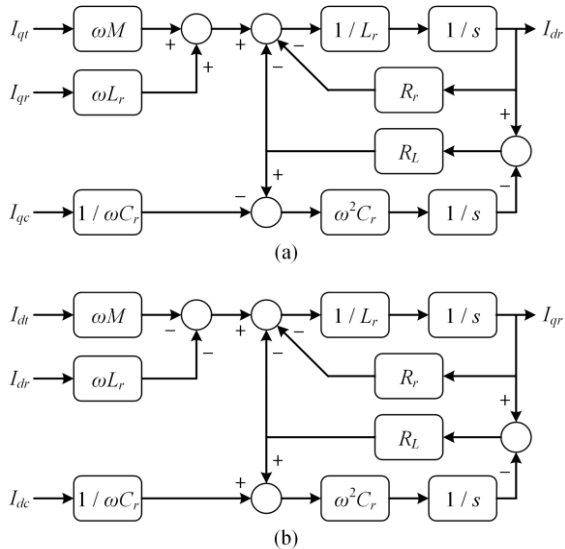

Fig. 4. Block diagrams of *dq* model on receiving side. (a) *d*-axis model. (b) *q*-axis model.

III. MUTUAL INDUCTANCE IDENTIFICATION

This section tries to design a mutual inductance identification scheme to show one *dq* model-based application in terms of measurement. According to the work [11], we could get the *d*-axis and *q*-axis currents from the measured *α*-axis and derived *β*-axis currents by *αβ*-to-*dq* transformation. Here we use the transmitting current as an example, and then the *αβ*-to-*dq* transformation is shown as

$$\begin{bmatrix} I_{dt} \\ I_{qt} \end{bmatrix} = \begin{bmatrix} \sin\theta & -\cos\theta \\ \cos\theta & \sin\theta \end{bmatrix} \begin{bmatrix} I_{\alpha t} \\ I_{\beta t} \end{bmatrix} \quad (16)$$

where $\theta$ is equal to $\omega t$.

Based on the *dq* model shown in (10), (11), and (12), a mutual inductance identification scheme implemented on the transmitter side is designed. Firstly, $I_{dt}$ and $I_{qt}$ can be got from (16), and the fundamental component of the input voltage can be calculated as

$$U_s = \frac{4}{\pi} U_{dc} \cos\left(\frac{\sigma}{2}\right) \quad (17)$$

where $\sigma$ is the phase-shift angle of the PWM drive waveform, and $U_{dc}$ is the input dc voltage of the full-bridge converter. Then, we divide (10) into the real part and the imaginary part and get the following equations.

$$-\omega M I_{qr} = U_s - R_t I_{dt} - \left(L_t + \frac{1}{\omega^2 C_t}\right)\frac{\Delta I_{dt}}{\Delta t} \\ + \left(\omega L_t - \frac{1}{\omega C_t}\right) I_{qt} \quad (18)$$

$$\omega M I_{dr} = -R_t I_{qt} - \left(L_t + \frac{1}{\omega^2 C_t}\right)\frac{\Delta I_{qt}}{\Delta t} \\ - \left(\omega L_t - \frac{1}{\omega C_t}\right) I_{dt} \quad (19)$$

Here the derivative parts are expressed in difference form by taking practical sampling into account. According to (14), $\omega M I_r$ can be calculated based on the known $\omega M I_{dr}$ and $\omega M I_{qr}$, as shown below

$$\omega M I_r = \sqrt{\left(-\omega M I_{qr}\right)^2 + \left(\omega M I_{dr}\right)^2} \quad (20)$$

Considering accumulating errors of deriving non-measured currents from measured ones [14], the derivatives of the currents on the receiving side should be ignored for (11) and (12). Then, $I_{dqc}$ in (11) can be eliminated with (12), and after multiplying $\omega M$ on both side of (11), the equation is shown as

$$\omega^2 M^2 I_t = \left| R_r + j\omega L_r + \frac{R_L}{1 + j\omega C_r R_L} \right| \omega M I_r \\ = |Z_r| \omega M I_r \quad (21)$$

As (14) said, the transmitting current amplitude $I_t$ can be calculated from the *d*-axis and *q*-axis currents. Then, using the known parameters, the mutual inductance can be obtained as

$$M = \sqrt{\frac{|Z_r| \omega M I_r}{\omega^2 I_t}} \quad (22)$$

So far, the *dq* model-based mutual inductance identification scheme is obtained.

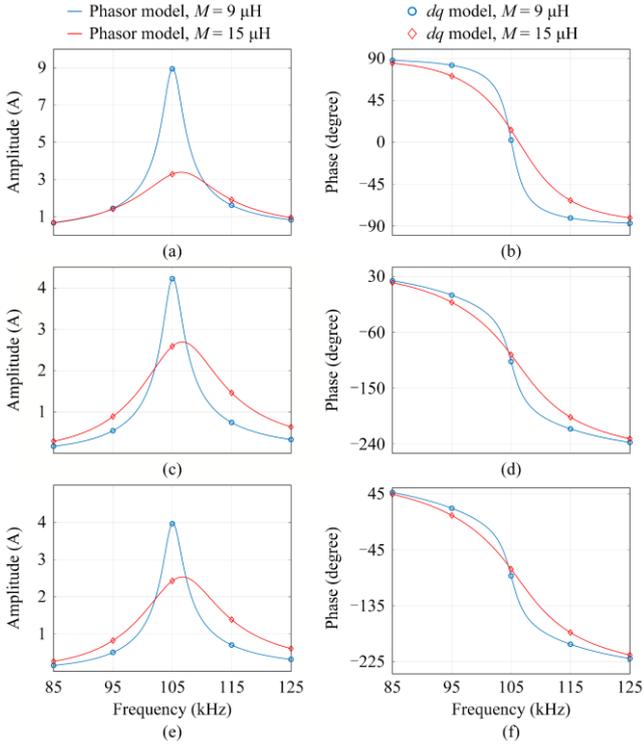

Fig. 5. Frequency responses of circuit phasor model and steady-state responses of $dq$ model. (a) Amplitude frequency property of $i_t$. (b) Phase frequency property of $i_t$. (c) Amplitude frequency property of $i_r$. (d) Phase frequency property of $i_r$. (e) Amplitude frequency property of $i_c$. (f) Phase frequency property of $i_c$.

## IV. VERIFICATION

This section presents the validation of the $dq$ model for SP compensated WPT systems as well as its one measurement application. The verification is conducted through numerical simulations in MATLAB/Simulink. The key parameters of the model are taken from [15], which are listed in Table I.

TABLE I. SYSTEM PARAMETERS

| Item | Value |
| --- | --- |
| Transmitting equivalent inductance ($L_t$) | 140.90 μH |
| Transmitting compensation capacitance ($C_t$) | 16.45 nF |
| Transmitting equivalent resistance ($R_t$) | 0.200 Ω |
| Receiving equivalent inductance ($L_r$) | 55.20 μH |
| Receiving compensation capacitance ($C_r$) | 41.47 nF |
| Receiving equivalent resistance ($R_r$) | 0.084 Ω |
| Equivalent load resistance ($R_L$) | 100 Ω |
| System operating frequency ($f$) | 105.0 kHz |
| Input dc voltage ($U_{dc}$) | 20 V |
| Mutual inductance ($M$) | Variable |

### A. Model Reliability

First, the frequency responses of the circuit of the SP compensated WPT systems are compared with the steady-state outputs from the developed $dq$ model. Here, the values of the mutual inductance $M$ are set to 9 and 15 μH in two groups of simulations. The lines shown in Fig. 5 are the frequency responses of the currents $i_t$, $i_r$, and $i_c$. Corresponding steady-state values of the $dq$ model are extracted from the Simulink model, which are marked with the circle and rhombus symbols on the plot. Fig. 5 shows the alignment between $dq$ model steady-state outputs and conventional phasor model frequency responses. It signifies the reliability of the proposed $dq$ model for SP compensated WPT systems.

Second, Fig. 6 shows the simulated waveforms when a step change from 10 V to 20 V occurs in the input dc voltage of the inverter. The voltage and current in the transmitting side are taken as an example. The figure incorporates the input voltage $u_s$, the transmitting current $i_t$, and the corresponding $d$-axis and $q$-axis currents. Notably, it is verified through the simulation results that the magnitude of $I_{dqt}$ is equal to the amplitude of $i_t$, namely (14).

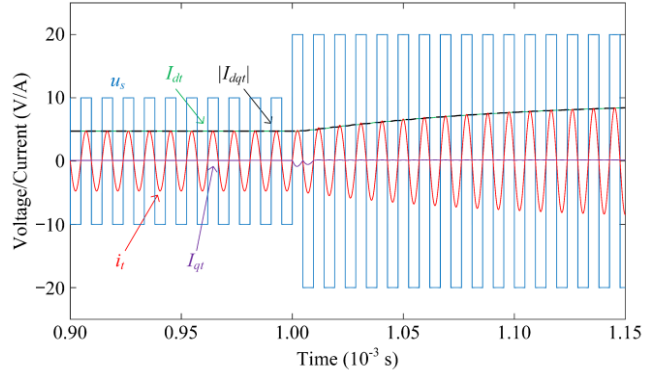

Fig. 6. Simulated waveforms under a step input dc voltage change in transmitting side.

Third, a set of simulations is carried out to determine if the initial phase angle calculation by $d$-axis and $q$-axis currents is correct. Similarly, the evaluations are done under different system operating frequencies, which is aimed to generate different initial phases. Table II shows the simulation results on the transmitting side. It is verified that the initial phase angle can be calculated from $d$-axis and $q$-axis current vectors according to (15).

TABLE II. VERIFICATION OF INITIAL PHASE ANGLE CALCULATION

| Operating frequency $f$ (kHz) | Actual $\varphi_t$ (degree) | Calculated $\varphi_t$ (degree) |
| --- | --- | --- |
| 85 | −88.044 | −88.036 |
| 95 | −83.180 | −83.182 |
| 105 | −1.229 | −1.232 |
| 115 | 81.911 | 81.914 |
| 125 | 86.819 | 86.820 |

The above three facets of the simulations effectively manifest the reliability of the proposed $dq$ model for SP compensated WPT systems.

### B. Evaluation of Mutual Inductance Identification

This subsection aims to evaluate the proposed $dq$ model-based mutual inductance identification scheme. The simulation results are taken from the Simulink model. Considering sampling frequencies of practical current sensors [14], the sampling frequency is set as 2520 kHz in the simulation, namely 24 samples per cycle.

First, two groups of simulations are run in two different load circumstances. Fig. 7 shows the results of the proposed identification method, and the relative errors are calculated to verify the accuracy. The coupling coefficient $k$ is set to change from 0.1 to 0.2 [1], and it is defined as

$$k = \frac{M}{\sqrt{L_t L_r}} \quad (23)$$

The selection of the scope of mutual inductance variation is mainly due to the demand of the normal WPT systems. The results got from the simulations prove that the relative errors of the identification method are mostly less than 3% under different loads of 50 and 100 Ω.

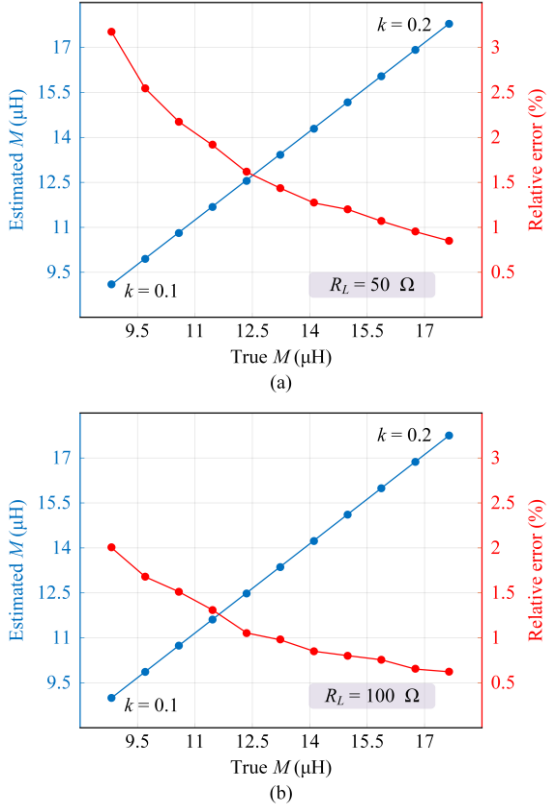

Fig. 7. Outcomes of mutual inductance identification. (a) 50 Ω load. (b) 100 Ω load.

As a complement, Table III shows the comparisons between the proposed mutual inductance identification scheme and others in recent years. It can be found that the relative error of the proposed method is less than 3%, which is smaller than the others. As a result, the proposed *dq* model-based mutual inductance identification scheme can be used in the SP compensated WPT-based applications.

TABLE III. COMPARISON OF ACCURACY

| Reference | Easy computation | Only on transmitting side | Relative error |
|---|---|---|---|
| [5] | ✓ | ✓ | 7.5% |
| [7] | ✗ | ✓ | 4% |
| [16] | ✓ | ✓ | 5.2% |
| This article | ✓ | ✓ | 3% |

## V. CONCLUSION

The development of WPT technology cannot be without the support of mathematical models. In this article, a *dq* modeling method for SP compensated WPT systems is proposed. As SP compensated WPT systems are widely used in specific scenarios like bioelectronics, the SP compensation network is selected as our research object. As derived, the proposed *dq* model can reflect the dynamic properties of the systems, thus assisting designs of measurement or control methods. Furthermore, a *dq* model-based mutual inductance identification method is presented to show the utility of the model. The simulation results from MATLAB/Simulink not only demonstrate the reliability of the *dq* model, but also quantify the error of the designed identification scheme, which is shown to have a relative error of less than 3%.

As a supplement, some future directions for *dq* model of SP compensated WPT systems may involve more tries on control method developments. It is a potential way forward to combine model predictive control, sliding mode control, or variable frequency control with the *dq* model, to further exploit the value of it.

REFERENCES

[1] K. Chen and Z. Zhang, "In-flight wireless charging: A promising application-oriented charging technique for drones," IEEE Ind. Electron. Mag., vol. 18, no. 1, pp. 6–16, March 2024.

[2] X. He, Y. Zeng, R. Liu, C. Lu, C. Rong, and M. Liu, "A dual-band coil array with novel high-order circuit compensation for shielding design in EV wireless charging system," IEEE Trans. Ind. Electron., vol. 71, no. 3, pp. 2545–2555, March 2024.

[3] W. Adepoju, I. Bhattacharya, M. Sanyaolu, and E. N. Esfahani, "Equivalent circuit modeling and experimental analysis of low frequency metamaterial for efficient wireless power transfer," IEEE Access, vol. 10, pp. 87962–87973, 2022.

[4] X. Qu, H. Han, S.-C. Wong, C. K. Tse, and W. Chen, "Hybrid IPT topologies with constant current or constant voltage output for battery charging applications," IEEE Trans. Power Electron., vol. 30, no. 11, pp. 6329–6337, November 2015.

[5] Y. Gu, J. Wang, Z. Liang, and Z. Zhang, "Mutual-inductance-dynamic-predicted constant current control of LCC-P compensation network for drone wireless in-flight charging," IEEE Trans. Ind. Electron., vol. 69, no. 12, pp. 12710–12719, December 2022.

[6] H. Pang, F. Xu, W. Liu, C. K. Tse, and K. T. Chau, "Impedance buffer-based reactance cancellation method for CLC-S compensated wireless power transfer," IEEE Trans. Ind. Electron., vol. 71, no. 7, pp. 6894–6906, July 2024.

[7] J. P.-W. Chow, H. S.-H. Chung, and C.-S. Cheng, "Use of transmitter-side electrical information to estimate mutual inductance and regulate receiver-side power in wireless inductive link," IEEE Trans. Power Electron., vol. 31, no. 9, pp. 6079–6091, September 2016.

[8] A. Laha and P. Jain, "Time domain modelling of a wireless power transfer system using a buck-boost converter for voltage regulation," 2021 IEEE Wireless Power Transfer Conference (WPTC), San Diego, CA, USA, 2021, pp. 1–4.

[9] H. Li, K. Wang, L. Huang, W. Chen, and X. Yang, "Dynamic modeling based on coupled modes for wireless power transfer systems," IEEE Trans. Power Electron., vol. 30, no. 11, pp. 6245–6253, November 2015.

[10] R. W. Erickson and D. Maksimovic, Fundamentals of Power Electronics. Norwell, MA, USA: Kluwer, 2020.

[11] W. Hong, S. Lee, and S.-H. Lee, "Sensorless control of series–series tuned inductive power transfer system," IEEE Trans. Ind. Electron., vol. 70, no. 10, pp. 10578–10587, October 2023.

[12] C. Jiang, K. T. Chau, C. Liu, and C. H. T. Lee, "An overview of resonant circuits for wireless power transfer," Energies, vol. 10, no. 7, June 2017.

[13] B. K. Kushwaha, G. Rituraj, P. Kumar, and P. Bauer, "Mathematical model for the analysis of series–parallel compensated wireless power transfer system for different misalignments," IET Circuits Devices Syst., vol. 13, no. 7, pp. 970–978, October 2019.

[14] K. Chen and Z. Zhang, "Rotating-coordinate-based mutual inductance estimation for drone in-flight wireless charging systems," IEEE Trans. Power Electron., vol. 38, no. 9, pp. 11685–11693, September 2023.

[15] M. Ishihara, K. Umetani, and E. Hiraki, "Strategy of topology selection based on quasi-duality between series–series and series-parallel topologies of resonant inductive coupling wireless power transfer systems," IEEE Trans. Power Electron., vol. 35, no. 7, pp. 6785–6798, July 2020.

[16] J. Vinotha, V. S. M. Amogh, M. P. Shreelakshmi, and R. R. Warier, "Mutual inductance estimation for series-parallel compensation topology in drone's wireless power transfer system," 2025 Fourth International Conference on Power, Control and Computing Technologies (ICPC2T), Raipur, India, 2025, pp. 1–6.